\documentclass[reprint, superscriptaddress, amsmath, amssymb, prl]{revtex4-2}
\usepackage{graphicx, ,siunitx}
\usepackage{multirow}

\usepackage{xspace}
\newcommand{\pg}{(p,$\gamma$)\xspace}
\newcommand{\pn}{(p,n)\xspace}
\newcommand{\MeVu}{MeV/nucleon\xspace}

\begin{document}
\newcommand{\gsi}{\affiliation{GSI Helmholtzzentrum f\"ur Schwerionenforschung GmbH, Darmstadt, Germany}}
\newcommand{\guf}{\affiliation{Goethe Universit\"at, Frankfurt am Main, Germany}}
\newcommand{\imp}{\affiliation{Institute of Modern Physics, Chinese Academy of Sciences, Lanzhou, China}}
\newcommand{\edin}{\affiliation{University of Edinburgh, Edinburgh, United Kingdom}}
\newcommand{\cenbg}{\affiliation{Université de Bordeaux, CNRS, LP2I Bordeaux, Gradignan, France}}
\newcommand{\jlu}{\affiliation{Justus-Liebig Universit\"at, Gie{\ss}en, Germany}}
\newcommand{\hij}{\affiliation{Helmholtz-Institut Jena, Jena, Germany}}
\newcommand{\herts}{\affiliation{Centre for Astrophysics Research, University of Hertfordshire, Hatfield, United Kingdom}}
\newcommand{\ptb}{\affiliation{Physikalisch-Technische Bundesanstalt, Braunschweig, Germany}}
\newcommand{\tud}{\affiliation{Technische Universit\"at Darmstadt, Darmstadt, Germany}}
\newcommand{\mpi}{\affiliation{Max-Planck-Institut f\"ur Kernphysik, Heidelberg, Germany}}
\newcommand{\tub}{\affiliation{Technische Universit\"at Braunschweig, Braunschweig, Germany}}
\newcommand{\aac}{\affiliation{University of Applied Sciences, Aachen, Germany}}
\newcommand{\triumf}{\affiliation{TRIUMF, Vancouver, British Columbia, Canada}}
\newcommand{\ubc}{\affiliation{Department of Physics and Astronomy, University of British Columbia, Vancouver, Canada}}
\newcommand{\uvic}{\affiliation{Department of Physics and Astronomy, University of Victoria, Victoria, British Columbia, Canada}}
\newcommand{\usai}{\affiliation{Department of Physics, Saitama University, Saitama, Japan}}
\newcommand{\utsu}{\affiliation{Tomonaga Center for the History of the Universe, University of Tsukuba, Ibaraki, Japan}}
\newcommand{\lanl}{\affiliation{Los Alamos National Laboratory, Los Alamos, NM, USA}}
\newcommand{\umichigan}{\affiliation{Facility for Rare Isotope Beams, Michigan State University, East Lansing, Michigan, USA}}
\newcommand{\ccindia}{\affiliation{Variable Energy Cyclotron Centre, 1/AF, Bidhan Nagar, Kolkata, India}}
\newcommand{\NCU}{\affiliation{Department of Physics, North Carolina State University, Raleigh, NC, USA}}
\newcommand{\triangleduke}{\affiliation{Triangle Universities Nuclear Laboratory, Duke University, Durham, NC, USA}}
\newcommand{\surrey}{\affiliation{School of Mathematics and Physics, University of Surrey, Guildford, United Kingdom}}
\newcommand{\unijena}{\affiliation{Institut für Optik und Quantenelektronik, Friedrich-Schiller-Universität Jena, Jena, Germany}}
\newcommand{\hfhf}{\affiliation{Helmholtz Forschungsakademie Hessen für FAIR (HFHF), Campus Gießen, Germany}}
\newcommand{\itastro}{\affiliation{Istituto Nazionale di Astrofisica, Osservatorio Astronomico d'Abruzzo, Teramo, Italy}}

\title{First Proton-Induced Cross Sections on a Stored Rare Ion Beam:\\ Measurement of $^{118}$Te(p,$\gamma$) for Explosive Nucleosynthesis}

\author{S.~F.~Dellmann}
\email{sdellmann@lanl.gov} \guf\lanl
\author{J.~Glorius} \gsi

\author{Yu.~A.~Litvinov} \gsi

\author{R.~Reifarth}
\guf\lanl
\author{L.~Varga}
\gsi
\edin
\author{M.~Aliotta}
\guf\gsi\edin
\author{F.~Amjad}
\gsi
\author{K.~Blaum}
\mpi
\author{L.~Bott}
\guf
\author{C.~Brandau}  
\gsi
\author{B.~Br\"uckner}
\guf
\author{C.G.~Bruno} 
\edin
\author{R.-J.~Chen}
\gsi\imp
\author{T.~ Davinson}
\edin
\author{T.~ Dickel}
\gsi\jlu
\author{I.~Dillmann}
\triumf\uvic
\author{D.~Dmytriev}
\gsi
\author{P.~Erbacher}
\guf
\author{O.~Forstner}
\gsi\hij
\author{D.~Freire-Fernández}
\lanl\mpi
\author{H.~Geissel}
\gsi\jlu
\author{K.~G\"{o}bel}
\gsi
\author{C.~J.~Griffin}
\triumf
\author{R.~E.~Grisenti}
\gsi
\author{A.~Gumberidze}
\gsi
\author{E.~Haettner}
\gsi
\author{S.~Hagmann}
\gsi
\author{T.~Heftrich}
\guf
\author{M.~Heil}
\gsi
\author{R.~Heß}
\gsi
\author{P.-M.~Hillenbrand}
\gsi
\author{C.~Hornung}
\gsi
\author{R.~Joseph}
\gsi
\author{B.~Jurado}
\cenbg
\author{E.~Kazanseva}
\gsi
\author{K.~Khasawneh}
\guf
\author{R.~Knöbel}
\gsi
\author{D.~Kostyleva} 
\gsi
\author{C.~Kozhuharov}  
\gsi
\author{I.~ Kulikov}  
\gsi
\author{N.~Kuzminchuk} 
\gsi
\author{D.~Kurtulgil}
\guf
\author{C.~Langer} 
\aac
\author{G.~Leckenby}
\triumf\ubc
\author{C.~Lederer-Woods}
\edin
\author{M.~Lestinsky}  
\gsi
\author{S.~Litvinov}  
\gsi
\author{B.~L\"oher}  
\gsi
\author{B.~Lorentz}  
\gsi
\author{E.~Lorenz}
\tud
\author{J.~Marsh} 
\edin
\author{E.~Menz}  
\gsi
\author{T.~Morgenroth}  
\gsi
\author{I.~Mukha}  
\gsi
\author{N.~Petridis}  
\gsi
\author{U.~Popp}  
\gsi
\author{A.~Psaltis}
\NCU\triangleduke
\author{S.~Purushothaman}
\gsi
\author{E.~Rocco}
\gsi
\author{P.~Roy}
\gsi\ccindia
\author{M.S.~Sanjari}  
\gsi\aac
\author{C.~Scheidenberger}  
\gsi\jlu\hfhf
\author{M.~Sguazzin}
\cenbg
\author{R.S.~Sidhu} 
\surrey
\author{U.~Spillmann} 
\gsi
\author{M.~Steck}  
\gsi
\author{T.~St\"{o}hlker}  
\gsi\hij\unijena
\author{A.~Surzhykov} 
\ptb\tub
\author{J.~A.~Swartz} 
\cenbg \umichigan
\author{Y.~Tanaka}  
\gsi
\author{H.~T\"ornqvist}
\gsi\tud
\author{D.~Vescovi}
\guf\itastro
\author{M.~Volknandt}
\guf
\author{H.~Weick}  
\gsi
\author{M.~Weigand}
\guf
\author{P.J.~Woods} 
\edin
\author{T.~Yamaguchi}  
\usai\utsu
\author{J.~Zhao}  
\gsi

\begin{abstract}
%
We present the first nuclear cross-section measurements of \pg and \pn reactions on $^{118}$Te at energies relevant for the $\gamma$-process nucleosynthesis. Absolute cross-section values for center-of-mass energies of 6, 7 and 10 MeV are provided, together with a theoretical extrapolation to the Gamow window. This experiment marks the first time that direct proton-induced reactions have been measured on a radioactive ion beam at the Experimental Storage Ring (ESR) at GSI, Darmstadt. 
This paves the way for a large variety of measurements, delivering new constraints for explosive nucleosynthesis and for physics beyond nuclear stability.

\end{abstract}
\maketitle

\textit{Introduction.}---Proton-capture reactions and their reverse reactions are important for our understanding of the origin of the elements. They are particularly relevant to the production of the so-called p-nuclei, a collection of stable, neutron-deficient isotopes whose synthesis cannot be explained by neutron capture processes. The $\gamma$ process, a large-scale nuclear conversion driven by the stellar explosion in different types of supernovae, is currently considered as their main production site \cite{PGR16, TRH18}. However, additional contributions, e.g. from the rp or $\nu$p processes, are the subject of ongoing debate~\cite{choplin2022, ScR06, FML06}.\\
Most of the nuclei involved in these $\gamma$-process reaction networks are radioactive. Despite the importance of knowing the reaction rates of these radioactive nuclei for reliable nucleosynthesis models, because of the huge experimental challenges, experimental data are very scarce. This introduces large uncertainties into the models and their predictions~\cite{PGR16, TRH18,ArG03,GGK15}.
By measuring the proton-capture reactions the important stellar ($\gamma$,p) reaction rates can be determined.  Deriving the stellar ($\gamma$,p) rates relevant to astrophysics from terrestrial (p,$\gamma$) cross sections is generally more reliable than from terrestrial $\gamma$-induced cross sections \cite{PhysRevLett.101.191101, Rau11}.\\
In this Letter, we present the first absolute cross section measurement of proton-induced reactions on the radionuclide $^{118}$Te with a half-life of 6.00(2) days~\cite{nndc}. The study is a key stage of the GSI proton-capture campaign at the Experimental Storage Ring (ESR) \cite{Fra87} and represents the first application of the inverse kinematics technique to a stored radioactive ion beam. Facilitated by recent improvements, we are able to provide absolute cross sections for the reactions $^{118}$Te\pg and $^{118}$Te\pn, which are part of the $\gamma$-process reaction network and can be used to probe nuclear reaction theory in unexplored regions of the nuclear chart. The energy range of the measurement covers 6, 7, and 10~MeV, which is just above the Gamow window for a stellar environment at \mbox{$T~=~3$~GK}. This is the temperature reached in the explosive environments that can produce the $\gamma$-process and as such it is interesting for nuclear astrophysics \cite{rauscher2010}. 

Inverse-kinematic techniques have been employed to directly study astrophysical proton-capture on radioactive beams since the early 1990s~\cite{Ruiz2014}. The first such study was conducted at Louvain-la-Neuve (Belgium), where the cross-section of $^{13}$N\pg was measured using in-beam $\gamma$ spectroscopy \cite{Decrock1991}. About a decade later, the first generation of recoil separators dedicated for nuclear astrophysics, specifically DRAGON at TRIUMF (Canada), yielded initial results with radioactive beams, focusing on lighter ions \cite{hutcheon2003}. Prominent examples are the measurements of $^{21}$Na\pg and $^{38}$K\pg \cite{bishop2003,Lotay2016}. In recent years, a few techniques and facilities targeting heavier ions have been developed. 
The storage ring technique has been developed with stable beam measurements of $^{96}$Ru(p,$\gamma$) and $^{124}$Xe(p,$\gamma$) at the ESR at GSI \cite{MAB15,GLS19}. The Separator for Capture Reactions (SECAR) recoil separator was constructed and is currently being commissioned to make use of radioactive ion beams with A~$\leq$~65 at FRIB (USA) \cite{berg2018}. Furthermore, the cross section of $^{83}$Rb\pg was measured using the EMMA spectrometer at TRIUMF \cite{Lotay2021}.

FRIB and GSI/FAIR rely on in-flight fragmentation for secondary beam production, which is a universal approach and covers a wide range of radioactive species, especially compared to the ISOL technique, which tends to deliver higher intensities but for a limited selection of secondary beams. 
Besides the general struggle for high intensity, the main challenge for astrophysical application of in-flight beams is to provide them in the low-energy regime of a few \MeVu after their production at relativistic energies.
At FRIB this is achieved by stopping and reaccelerating beams \cite{vbh23}. At GSI secondary beams can be stored and decelerated in the ESR, which also offers a variety of additional beam manipulations, as described below \cite{GLORIUS2023190}.

\begin{figure}[t]
\includegraphics[width=\linewidth]{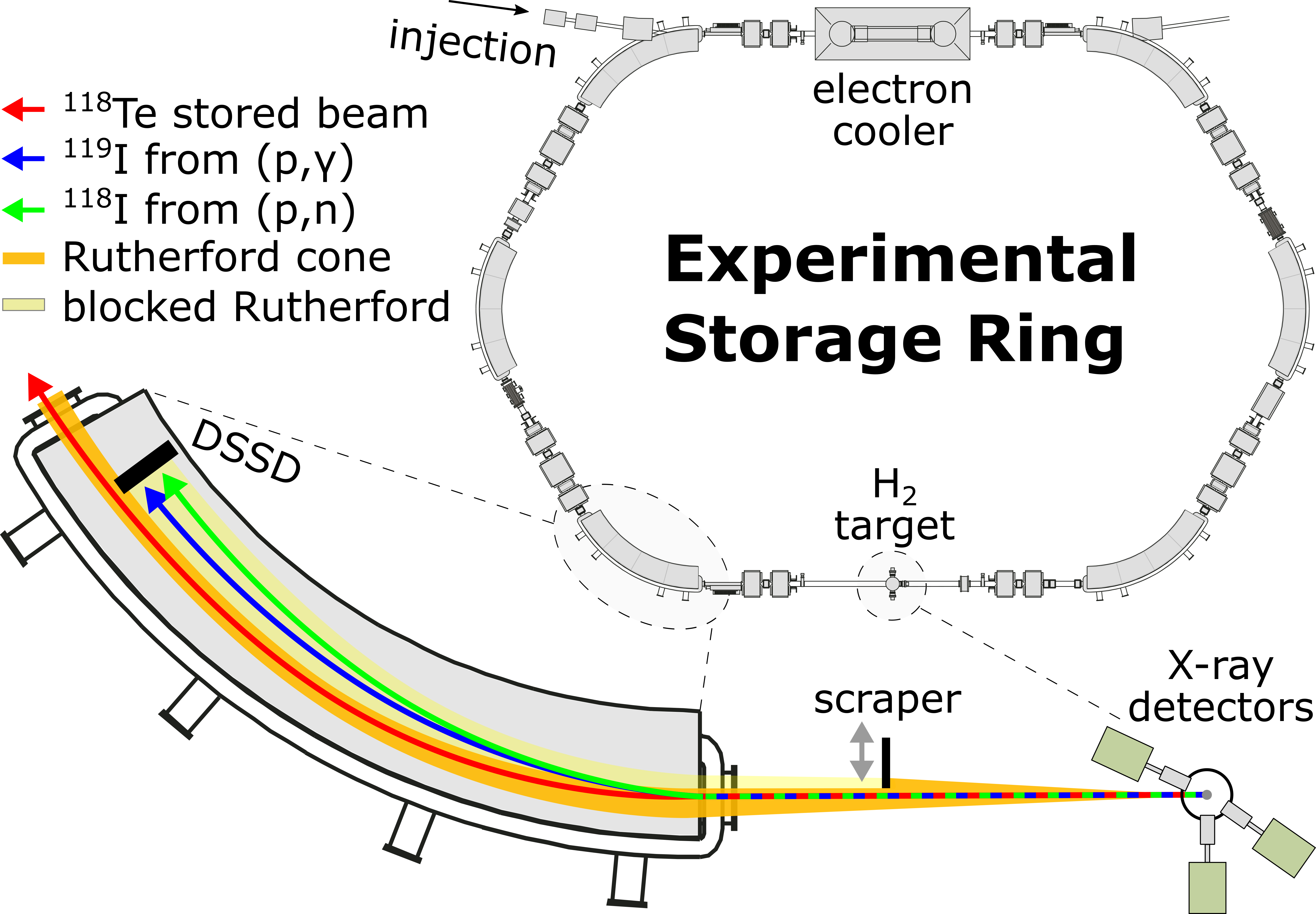}
\caption{\label{fig:seperation} Overview drawing of the ESR with a zoom on the internal gas-jet target and recoil detection areas. The separation of the reaction products inside the dipole magnet is shown schematically. The \pg and \pn recoil ions are intercepted by the Si detector at the end of the dipole, while the ERASE scraper blocks most of the Rutherford scattering background directly before the magnet entrance. HPGe detectors surround the target area to measure the characteristic X-ray emission of the atomic electron-capture process used for normalization of the nuclear cross sections.}
\end{figure}

\textit{Experiment.}---The cross section measurement of $^{118}$Te(p,$\gamma$) and $^{118}$Te(p,n) was performed at GSI. By the unique combination of the Fragment Separator~(FRS)~\cite{GAB92} and the ESR~\cite{Fra87}, the production and storage of the radioactive isotope $^{118}$Te was accomplished. A stable $^{124}$Xe beam from the UNIversal Linear ACcelerator (UNILAC) was accelerated to about 550~\MeVu in the SchwerIonenSynchrotron~(SIS18) and extracted to impinge on the 2.5~g/cm$^2$ Be production target at the entrance of the FRS. After an additional stripper stage, fragments of $^{118}$Te with a charge state of $q=52^+$ were filtered out and injected into the ESR at 400~\MeVu. In the ring, stochastic cooling \cite{NOLDEN2004329} and beam accumulation \cite{nolden2013} of several injections was applied in order to increase the stored intensity to about $7 \times 10^6$ ions. Subsequently, the ions were decelerated to low energies and subjected to permanent electron cooling \cite{steck2004} in order to provide a beam of small momentum spread $\Delta p/p \sim 10^{-5}$ and also to compensate for energy loss in the H$_2$ target \cite{KPW09}. At this point, about $10^6$ ions of bare $^{118}$Te$^{52+}$ with a storage life-time of about 1.5~s were collided with the H$_2$ target to trigger p-induced reactions. With an ion revolution frequency of 300 to 400~kHz and a target density of about 10$^{14}$~atoms/cm$^2$, peak luminosities above 10$^{25}$~cm$^{-2}$s$^{-1}$ were reached. The entire beam preparation procedure is described in more detail in \cite{GLORIUS2023190}.

The first dipole magnet of the ring situated downstream from the target provides magnetic separation between the reaction products and the stored beam, as the charge of the (p,$\gamma)$ and (p,n) recoil ions is increased by one, while the momentum remains close to the one of the stored beam. Figure~\ref{fig:seperation} shows schematically the separation within the dipole magnet. The fully stripped $^{119}$I$^{53^+}$ recoils (blue line) from (p,$\gamma$) are shifted by about 48~mm to an inner orbit with respect to the stored beam, while $^{118}$I$^{53^+}$ recoils (green line) from (p,n) experience a stronger deflection by about 70~mm at the end of the magnet. At this position, a Micron W1-type Double-sided-Silicon-Strip Detector (DSSD) \cite{micron,sellin1992} was used, providing a position resolution of 3~mm by means of individual readout of the $16\times16$ strips in horizontal and vertical orientations, respectively. 

In the precursor experiments a strong background distribution from elastic Rutherford scattering at the H$_2$ target covered the entire detector area and limited the sensitivity of the method. To avoid this, a new scraper system has been introduced, which improves the signal-to-background ratio for nuclear recoil signatures on the DSSD by about one order of magnitude. It was instrumental in making the radioactive ion beam experiment possible in the first place \cite{VARGA2024submitted}.

At the target region three planar High-Purity Germanium~(HPGe) detectors were employed to measure the X-ray emission at angles of 35\textdegree, 90\textdegree, and 145\textdegree. These detectors enable a measurement of the nuclear reaction cross section relative to the process of radiative electron capture from the H$_2$ target into the K-shell of the bare stored ion, the so-called K-REC process, for which the cross sections were calculated theoretically \cite{EiS07}. 

It has to be noted that the $^{118}$Te(p,n) reaction channel has a Q-value of $Q_{\text{(p,n)}}=-7.50(3)$~MeV and is accordingly open only for the highest energy investigated here at $E_{\text{CM}}$ = 10.05 MeV.

\textit{Analysis.}---The cross section of the proton-induced reactions based on a luminosity normalization via the \mbox{K-REC} process is calculated using the formula :

\begin{equation}
     \sigma_{(\text{p,x})} = \frac{N_{(\text{p,x})}}{N_{\text{KREC}}} \epsilon_{\text{KREC}} \Delta \Omega \frac{d\sigma_{\text{KREC}}}{d\Omega}.
\end{equation}

Here, $N_{(\text{p,x})}$ is the number of detected recoil ions of the respective nuclear reaction channel (x = $\gamma$, n) in the DSSD, and $N_{\text{KREC}}$ is the number of counts registered in the K-REC peak detected with efficiency $\epsilon_{\text{KREC}}$ within the solid angle $\Delta \Omega$ of a HPGe detector. The differential cross section $\frac{d\sigma_{\text{KREC}}}{d\Omega}$ was theoretically determined within the framework of relativistic impulse approximation, see Ref. \cite{EiS07}  for further details. The uncertainty of these K-REC calculations is about $1\%$ as was estimated by using different models of the target electron momentum distributions. 

The recorded X-ray spectra show a unique pattern for each elemental species stored in the ring \cite{ZGO2022}. Thus, additionally to the luminosity normalization via K-REC, the spectra are also used to examine the purity of the stored fragment beam. According to simulations using LISE++~\cite{TARASOV20084657}, small amounts of $^{123}$Xe$^{54+}$, $^{120}$I$^{53+}$ or $^{116}$Sb$^{51+}$ ions were expected to be injected and stored in the ESR \cite{TaB08}. While these species were visible during the accumulation phase, the subsequent radio frequency~(RF) modulation needed for deceleration of the beam is very selective and acts as an effective contamination filter. The X-ray spectra provide proof of a relative purity of the stored $^{118}$Te beam to a level of 10$^{-2}$ or better, see~\cite{GLORIUS2023190} for details.

The nuclear reaction products of $^{118}$Te+p were recorded using the DSSD inside of the dipole magnet. For heavy ion hits in the present energy range of 0.7 to 1.2~GeV, the DSSD provides a detection efficiency of 100$\%$. Spatial information about the ion hits is obtained by coincidence conditions between front- and backside strips of the detector, which create $3 \times 3$~mm$^2$ pixels in their respective overlap regions due to their perpendicular orientation.

In order to cleanly identify ion hits and their origin, the DSSD spectra were compared to Monte Carlo simulations using the code MOCADI for ion transport and two-body kinematics \cite{IWG11}. Further details about the use of MOCADI can be found in Ref.~\cite{XING2020164367, Dellmann2024}. For the full analysis an internal energy calibration \cite{RGG15} and a consideration of detector response features of the DSSD had to be taken into account. For instance, so-called inter-strip hits, when an ion deposits energy in the gap region between adjacent strips, had to be filtered out, in order to get clean ion energy spectra. This enabled the application of additional conditions in terms of ion energy and position to further reduce the background.

\begin{figure}[t!]
    \centering
    \includegraphics[width=\linewidth]{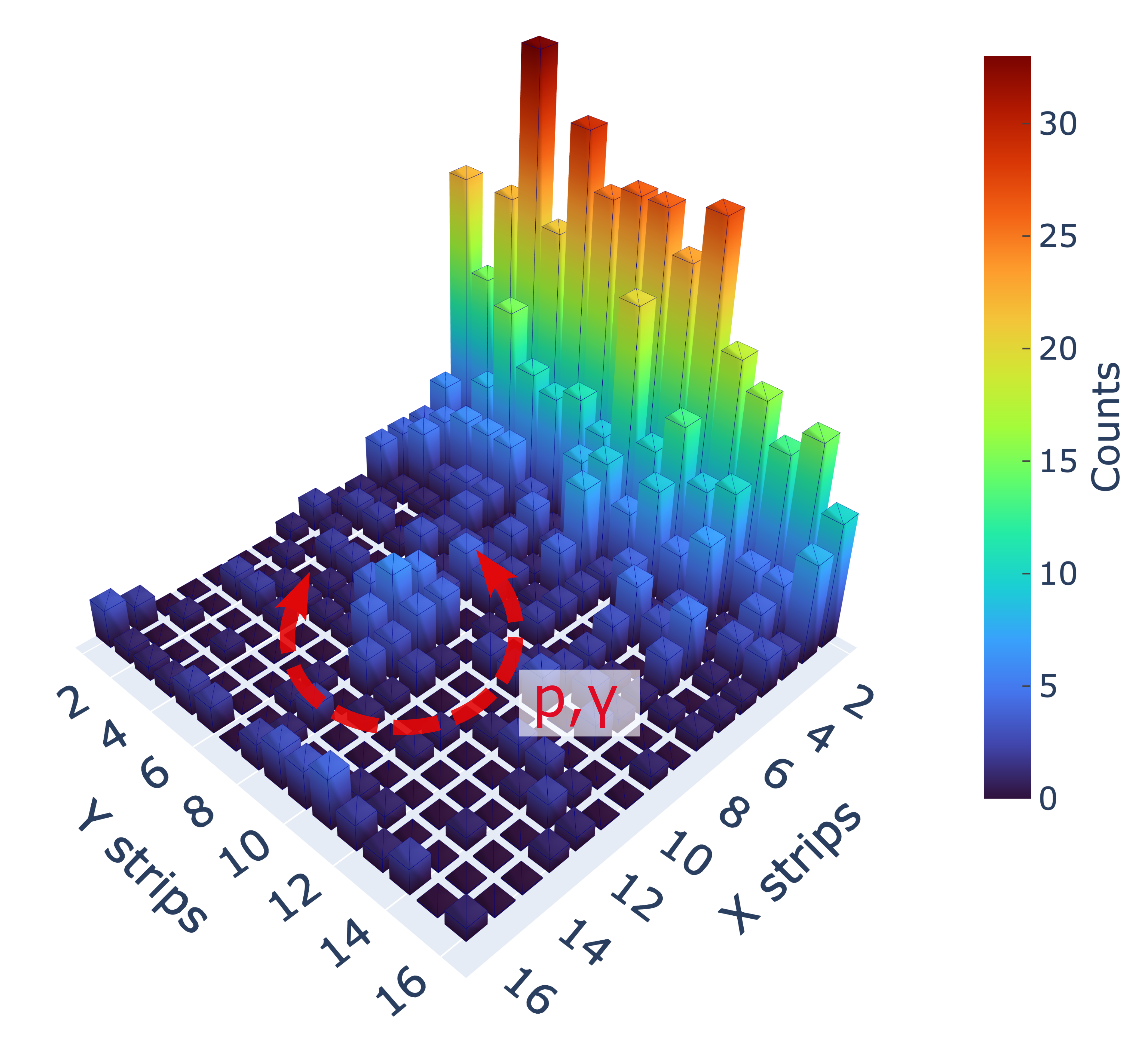}
    \caption{\label{fig:DSSD} Ion-hit distribution recorded with the DSSD for a measurement time of 52~h with $^{118}$Te ions stored at 6~\MeVu. The central bump within the red contour originates from the $^{118}$Te\pg reaction, while $^{118}$Te ions elastically scattering off the hydrogen target atoms produce the background distribution on the upper right. The ERASE blocking removes the dominant part of this Rutherford cone below the \pg peak, but a weak tail, presumably from secondary scattering off the scraper edge, still extends into this region.}
\end{figure}

Figure~\ref{fig:DSSD} shows the raw ion-hit distribution for $^{118}$Te(p,$\gamma$) measured over 52 hours at 6 MeV/nucleon. A narrow cluster of (p,$\gamma$) recoil ions is visible on top of the strongly suppressed Rutherford background. The edge of the detector experiencing higher background rates is about 25~mm away from the stored beam. The blocking scraper located before the dipole magnet is positioned about 23~mm from to the beam for this dataset. 

To determine the number of \pg events within a measurement, the corresponding region of the hit distribution is integrated and the residual background is fitted and subtracted. This procedure was done with a 2-dimensional histogram (Fig.~\ref{fig:DSSD}), as well as for a 1-dimensional projection, for which the background was described by an exponential function plus a first degree polynomial. Both approaches agree well within the uncertainties. Further details about the different approaches on the background determination can be found in references \cite{Dellmann2024} and \cite{Dellmann2023}.  For the (p,n) data at 10.05~MeV, an additional efficiency $\epsilon_{\text{(p,n)}}~=~64(12)\%$ had to be evaluated based on the simulations \cite{Var20}, because the (p,n) distribution at this energy extends beyond the active area of the DSSD. 

\begin{table}[b!]
\caption{\label{tab:table4}%
Cross-sections results for the \pg and \pn reactions on $^{118}$Te.} 
\begin{tabular*}{0.9\columnwidth}{@{\extracolsep{\fill}} l c c}
\hline\hline
 & \textbf{$E_{\text{CM}}$} & \textbf{$\sigma$}  \\
\multirow{-2}{*}{\textbf{Reaction}} & \textbf{[MeV]} & \textbf{[mbarn]} \\ \hline \rule{0pt}{2.4ex}
$^{118}$Te(p,$\gamma$) &  \hphantom{1}6.04  $\pm$ 0.02                 & \hphantom{1}44.4 $\pm$ \hphantom{1}4.4$_{\text{stat}}$ $\pm$ \hphantom{1}5.6$_{\text{sys}}$  \\ \hline \rule{0pt}{2.4ex}
$^{118}$Te(p,$\gamma$) &  \hphantom{1}7.05    $\pm$ 0.02                  & \hphantom{1}71.4 $\pm$ \hphantom{1}5.0$_{\text{stat}}$ $\pm$ \hphantom{1}3.8$_{\text{sys}}$  \\ \hline \rule{0pt}{2.4ex}
$^{118}$Te(p,$\gamma$) &                         & \hphantom{1}35.7 $\pm$            12.5$_{\text{stat}}$ $\pm$ \hphantom{1}5.4$_{\text{sys}}$   \\ \rule{0pt}{2.4ex}
$^{118}$Te(p,n)        & \multirow{-2}{*}{10.05 $\pm$ 0.02}       &            403.2 $\pm$            36.8$_{\text{stat}}$ $\pm$            88.0$_{\text{sys}}$  \\ \hline\hline
\end{tabular*}
\end{table}

The cross section values for $^{118}$Te(p,$\gamma$) and $^{118}$Te(p,n) reactions are listed in Table~\ref{tab:table4} along with their uncertainties. The statistical uncertainties are determined from Poisson statistics before background subtraction, while the systematic uncertainties have multiple origins. The major systematic contribution is the description of the background, it strongly depends on the signal-to-background ratio in the respective measurement  ($\pm$ 5-13\%). Further systematic components are the K-REC cross section uncertainties ($\pm$  1\%) as well as the detection efficiency for X-rays ($\pm$ 5-6\%). The efficiency correction for interstrip events contributes with an uncertainty of about ({$\pm$} 2\%). 
Finally, for the case of the (p,n) analysis the dominating uncertainty is the determination of the geometric detection efficiency. Based on variations of all conceivable parameters regarding the shape of the (p,n) distribution and the geometry between the DSSD and the beam, a conservative uncertainty of {$\pm$} 20\% has been adopted.

\begin{figure}[t]
    \centering
    \includegraphics[trim={1cm 0 3cm 1cm},clip,width=\linewidth]{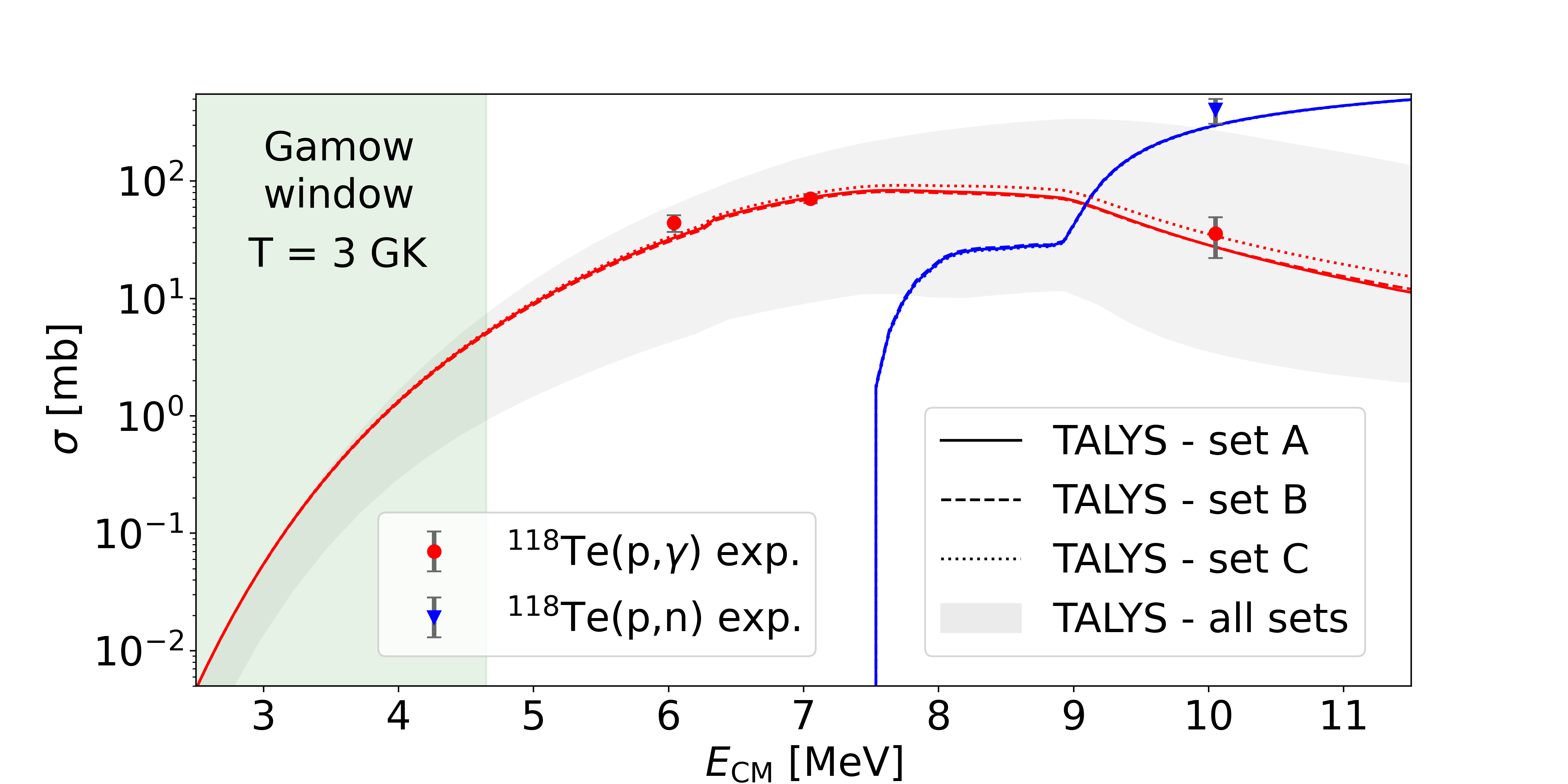}
    \caption{Cross sections of $^{118}$Te\pg in red and $^{118}$Te\pn in blue are depicted for the measurement region and the Gamow window at $T = 3$~GK. The experimental data is shown in overlay with TALYS calculations based on the three parameter sets that generate the best reproduction. The gray band represents the full parameter space available in TALYS without individual tuning of the underlying models.}
    \label{fig:WQ}
\end{figure}

\textit{Discussion}---The cross section results are illustrated in Fig.~\ref{fig:WQ}. To apply these \pg data to astrophysical scenarios an extrapolation to the Gamow window for an appropriate temperature is needed. For the $\gamma$ process an energy range of 2.4--4.9~MeV is relevant for an estimated temperature of $T = 3$~GK \cite{rauscher2010}. To this end, the experimental cross sections are shown in comparison to theoretical predictions performed with TALYS~1.96~\cite{Koning2007}. These calculations are based on the Hauser-Feshbach formalism and different models for the underlying physics parameters~\cite{HaF52}. 

The options to choose physics models in TALYS~1.96 for description of the optical-model potential (OMP), the nuclear level density (NLD) and the $E1$ and $M1$ photon strength function (PSF), are manifold. In order to find the best reproduction of the experimental data, a chi-square analysis was conducted with a focus on the low-energy extrapolation of the \pg channel. 

In total, three parameter sets could be identified with $\chi^2_{\text{red}}$ very close to unity for the full dataset. These theoretical cross sections are shown in overlay with the experimental data in Fig.~\ref{fig:WQ}. 

All three sets, called A, B \& C here, are based on the JLM-B optical potential (TALYS keyword: \mbox{{\it jlmomp~y}}) obtained with the Skyrme Hartree-Fock-Bogoliubov (HFB) matter densities ({\it radialmodel~1}) and a modified normalization coefficient for the imaginary part  \mbox{({\it jlmmode~3})}~\cite{PhysRevC.88.024308}. Moreover, all sets utilize the 
Back-Shifted Fermi Gas model ({\it ldmodel~2})~\cite{IST75} for description of the level density. While a few other combinations of OMP and NLD parameters can reproduce the \pg channel on an acceptable level, none of them are able to reproduce the \pn data point at the same time.

The PSF model selection is not as crucial and is the only difference between the three highlighted sets here. We use Goriely's hybrid model ({\it strength~5})~\cite{goriely98} for set A, the Simple Modified Lorentzian model ({\it strength~9})~\cite{Plujko2019,PhysRevC.99.014303} for set B and the D1M-Gogny HFB model ({\it strength~8})~\cite{PhysRevC.98.014327} for set C.

While set A produces the best $\chi^2_{\text{red}}(A) = 1.015$, the results with set B also deliver a very good reproduction of the experimental data with $\chi^2_{\text{red}}(B) = 1.052$. Set C gives $\chi^2_{\text{red}}(C) = 1.131$ and a slightly higher prediction of the \pg cross section in the experimental energy range. All three calculations reproduce the full dataset within a $2\sigma$ environment of the experimental error bars.

Based on our chi-square analysis, we recommend set A for low-energy application in TALYS~1.96 when dealing with $^{118}$Te(p,$\gamma$) and $^{118}$Te(p,n). A reaction rate for \mbox{$T = 3$~GK} is calculated by TALYS as \mbox{$\lambda_{\text{\pg}}(A) = 21.94$~s$^{-1}$}. The other two sets generate deviations from this rate of only $1\%$ for set B and $7\%$ for set C. 

Other parameter sets in TALYS, which reproduce the experimental \pg data well, but not the \pn data point, partly cause large deviations for the extrapolation and rate calculation.
Here, the simultaneous  measurement of \pg and \pn demonstrates its strength, enabling the application of much stronger constraints on the Hauser-Feshbach parameter selection. 

While impossible to quantify, it is obvious that the uncertainty of the reaction rate for $^{118}$Te(p,$\gamma$) is reduced tremendously as compared to the situation before this measurement. This is illustrated by inspecting the full parameter space available in TALYS~1.96 as indicated by the gray band in Fig.~\ref{fig:WQ}, which spans up to one order of magnitude in \pg cross section and causes even larger uncertainties in rate calculation.

The approach followed here determines the best performing sets of Hauser-Feshbach parameters for low-energy extrapolation of a cross section to be used for reaction rate calculation in astrophysical scenarios. For a systematic investigation of the underlying physics models one might choose a different strategy, which might also consider data on nuclei in the close vicinity or on a more global scale. While this is of major importance in moving towards a global description of nuclear reactions in unexplored regions of the nuclear chart, it remains beyond the scope of the present manuscript. But it shall be perceived as a strong motivation for further reaction studies with exotic systems. 


\textit{Conclusions}---With this experiment it could be demonstrated for the first time that a simultaneous  measurement of \pg and \pn reactions is feasible with stored radioactive ions. The measurements on $^{118}$Te were performed close to the Gamow window of explosive nucleosynthesis in supernovae or X-ray bursts. The extrapolation towards lower energies is based on Hauser-Feshbach calculations using optimized parameters.

The simultaneous  measurement of the \pg and \pn reaction channels resulted in strong constraints for the nuclear model selection essential to the theoretical description of the cross section.

This experimental technique is now established for rare ion beams and shall encourage future measurements with even more exotic beams. Future campaigns will aim to measure challenging key reactions of explosive nucleosynthesis, such as $^{59}$Cu\pg in the rp process or $^{91}$Nb\pg in the $\gamma$ process \cite{meisel2019,RNH16}. With the new CRYRING facility at GSI, used as a low-energy extension of the FRS-ESR complex, even lower energies directly inside the Gamow window are within reach~\cite{LAA16,marsh2024}. Knowledge about the experimental values of these reactions will build a grid that will have an impact on theoretical predictions in nucleosynthesis studies. It may also provide clues to better understand the underlying nuclear structure. The astrophysical influence of the reactions measured here is the subject of further investigations \\

\begin{acknowledgments}
\textit{Acknowledgments}---This project has received funding from the European
Research Council (ERC) under the European Union’s
Horizon 2020 research and innovation programme (grant agreement No 682841 “ASTRUm”). This work is further supported by the European Union (ChETEC-INFRA, project no. 101008324), the Helmholtz Forschungs-Akademie Hessen for FAIR (HFHF), the Federal Ministry of Education and Research (BMBF) under Grant No.  05P19RFFA2, 05P15RFFAA and 05P15RGFAA and the State of Hesse within the Research Cluster ELEMENTS (Project ID 500/10.006). BJ, MS and JS acknowledge support
from the ERC grant No 884715, NECTAR. AP acknowledges support from the Deutsche Forschungsgemeinschaft (DFG, German Research Foundation) Project No.279384907 SFB 1245 and the European Research Council grant No. 677912 EUROPIUM. ID, CG, and GL acknowledge funding from the Natural Sciences and Engineering Research Council of Canada (NSERC, SAPIN-2019-00030). MA acknowledges support from the EMMI Visiting Professorship scheme. RSS and PJW acknowledge support from the Science and Technology Facilities Council (STFC) (Grant No. ST/P004008/1). CLW acknowledges support from the European Research Council grant No. 677497, DoRES
\end{acknowledgments}

%

\end{document}